\documentclass[useAMS,usenatbib]{mn2e}
\usepackage{graphicx}
\usepackage{epsfig}

\title[{\it Suzaku} observations of G296.1$-$0.5]{Ejecta detection in the middle-aged Galactic supernova remnant G296.1$-$0.5 observed with {\it Suzaku}}
\author[F.~G\"{o}k, A. Sezer]{F.~G\"{o}k,$^{1}$\thanks{E-mail: gok@akdeniz.edu.tr
(FG); aytap.sezer@uzay.tubitak.gov.tr (AS)} A. Sezer $^{2,3}$
\footnotemark[1]\thanks{This file has been amended to highlight
the proper use of \LaTeXe\ code with the class file. These changes
are for illustrative purposes and do not reflect the
original paper by F.~G\"{o}k.}\\
$^{1}$Akdeniz University, Faculty of Sciences, Department
of Physics, Antalya, 07058, Turkey\\
$^{2}$T\"UB\.ITAK Space Technologies Research Institute, ODTU
Campus, Ankara, 06531, Turkey\\
$^{3}$Bo\~gazi\d{c}i University, Faculty of Art and Sciences,
Department of Physics, \.Istanbul, 34342, Turkey\\
}
\begin{document}

\date{}

\pagerange{\pageref{firstpage}--\pageref{lastpage}} \pubyear{2011}

\maketitle

\label{firstpage}

\begin{abstract}
In this paper, we report the detection of ejecta in the
middle-aged Galactic supernova remnant G296.1$-$0.5 with the X-ray
Imaging Spectrometer onboard the {\it Suzaku} satellite. The
spectra of three lobes, north, southeast and southwest and
inter-lobe regions, consist of soft (0.3$-$2.0 keV) emission
originated from non-equilibrium ionization plasma. In north,
southeast and inter-lobe regions, the thermal emission can be
represented by a one-component, in southwest region it can be
represented by two-component non-equilibrium ionization (VNEI)
model. The spectra of studied regions have lines of N, O, Ne, Mg
and Si elements. Si emission from this remnant is shown for the
first time in this work. Enhanced abundances of Ne, Mg and Si
elements obtained show the ejecta contribution in all regions.
Assuming that the remnant is in Sedov phase, we obtained ambient
density $n_{0}\sim 0.45$ cm$^{-3}$, age $t$ $\sim 2.8\times10^4$
yr, shock velocity $V_{\rm s}\sim 320$ km s$^{-1}$, shock
temperature $T_{\rm s}\sim 1.2\times10^{6}$ K, and swept-up mass
$M_{\rm sw}\sim 340$ {M\sun} at an adopted distance of d=3 kpc.
\end{abstract}

\begin{keywords}
ISM: supernova remnants$-$ISM:
individual(G296.1$-$0.5)$-$X-rays:ISM
\end{keywords}

\section{Introduction}

The shell type Galactic supernova remnant (SNR) G296.1$-$0.5
($\rmn{RA}(2000)=11^{\rmn{h}} 51^{\rmn{m}} 10^{\rmn{s}}$,
$\rmn{Dec.}~(2000)=-62\degr 34\arcmin$) with 37$\times$25
arcmin$^{2}$ angular size was discovered in radio band at 408 MHz
\citep{b24,b5}. \citet{b21} performed optical observation of the
remnant and reported the identification of the nebulosity and its
filamentary nature. \citet{b37} suggested that there was another
SNR located at the northeast part of the radio remnant. Since the
region in which the remnant is located on the Galactic plane is
rich in SNRs, this suggestion seemed reasonable. \citet{b38}, with
an improved radio map at 408 MHz, and \citet{b3}, with X-ray data
from {\it EXOSAT} and {\it Einstein} observations, reported that
this is a single, complex remnant. From {\it ROSAT} X-ray data
\citet{b39} obtained an interstellar absorbtion of $N_{\rm
H}\geq1.5\times10^{21}$ cm$^{-2}$ and a low electron temperature
$kT_{\rm e}\sim0.2$ keV from single temperature model fitting.
They also applied two temperature component model, obtained
electron temperatures of $\sim$0.1 keV and $\sim$3.5 keV for each
component. Assuming the distance of 4 kpc and ellipsoidal shell
geometry for the remnant, they found total X-ray emitting mass and
ambient density to be $\sim$250 {M\sun} and $\sim$0.8 cm$^{-3}$,
respectively. Using the Sedov equations, they found an age of
$\sim$$2\times10^{4}$ yr. From the data of {\it XMM-Newton}
observation, \citet{b40} found the electron temperature $kT_{\rm
e}$=0.6 keV by applying a non-equilibrium ionization model (VNEI)
to the spectra. They obtained a low absorbing column density,
$N_{\rm H}=(2-4)\times10^{20}$ cm$^{-2}$, excess of N, and a
deficit of O. They suggested a massive progenitor star and
concluded that the remnant was resulted from core-collapse
supernova. They have detected a transient source 2XMMi
J115004.8–622442 at the edge of this remnant.

Various distance estimates were given from different studies;
\citet{b21} estimated the distance d=$3\pm1$ kpc by using
H$\alpha$ and H$\beta$ observations and reddening measurements in
the direction of crux \citep{b46}, and by fitting the radial
velocity ($-$35 km s$^{-1}$ with respect to the local standard of
rest) to galactic rotation. From $\Sigma-$D relation \citet{b38},
\citet{b22} and \citet{b24} adopted d=7.7 kpc, d=6.6 kpc and d=4.9
kpc, respectively. \citet{b40} assumed d=2 kpc, considering the
giant HI shell, GSH 304-00-0.5 located in front of the remnant at
d $\sim$ 1.2 kpc.

Our aim is to study middle-aged SNRs to investigate interstellar
medium (ISM)/shock interactions and ISM itself, i.e., the status
of the ionization state, temperature (line emission) variation all
over the remnant. G296.1$-$0.5 is particularly interesting one
with its complex shock and plasma structure. The Japanese X-ray
observatory {\it Suzaku} \citep{b17}, since it has a large
collecting area and low background, is capable of resolving line
emission characteristic of SNRs, especially at low energies, that
provides valuable information in understanding SNRs. By using the
archival data of {\it Suzaku}, we were able to produce higher
quality image and the spectra of the remnant, which motivated us
for this study.

This paper is organized as follows. We describe the observations
and the data reduction in Section 2. In Section 3, we explain the
details of image and spectral analysis. In Section 4, we discuss
our results and description of the X-ray emission from
G296.1$-$0.5.

\section[]{Observations and Data Reduction}

The {\it Suzaku} observations of the north and south regions of
G296.1$-$0.5 were made on 2007 August 09 and on 2008 January 17,
respectively, with X-ray Imaging Spectrometer (XIS: \citet{b12}).
The observation ID and exposure time are 502068010, 502069010 and
77.2, 69.2 ksec, respectively. The XIS consists of four sets of
X-ray CCD camera system (XIS0, 1, 2, and 3). XIS1 has a
back-illuminated (BI) sensor, while XIS0, 2, and 3 have
front-illuminated (FI) sensors. The XIS2 sensor was available only
until 2006, therefore, we use data of XIS0, XIS1, XIS3. The XIS
has a field of view (FOV) of a 17.8$\times$17.8 arcmin$^{2}$.

In both observations, the XIS was operated in the normal
full-frame clocking mode with the standard $3\times3$ and
$5\times5$ editing mode. Response matrices and ancillary response
files (ARFs) were generated for each XIS independently using {\sc
xissimrmfgen} version 2007-05-14 and {\sc xissimarfgen} version
2008-04-05 \citep{b7}. Reduction and analysis of the data were
performed following the standard procedure using the {\sc headas}
v6.4 software package, and spectral fitting was performed with
{\sc xspec} v.11.3.2 \citep{b2}.

\section{Analysis}
\subsection{Image Analysis}

Figure 1a and 1b present the smoothed XIS0 images of north and
south regions of G296.1$-$0.5 in 0.3$-$10 keV energy band,
respectively. In  the northern part there is one bright lobe,
while in the southern part there are two bright lobes. So, to
determine the temperature variations and ionization states, we
studied the bright lobes and a region between the lobes in the
south part as shown in Fig. 1a and 1b with solid ellipses. The
regions are abbreviated as north (N), southeast (SE), southwest
(SW) and inter-lobe (M) region. M region is selected as the
reference location in order to highlight the peculiar properties
of the lobes if there is any. Dashed ellipses in both figures show
the regions chosen for background subtraction. The calibration
sources located at the two corners in the FOV are excluded.

Figure 2 gives the mosaic image of XIS0 in 0.3$-$10 keV energy
band, N and S, which are overlaid with the radio image obtained at
843 MHz by \citet{b25} for comparison, the positions of optical
filaments \citep{b21}, H${\alpha}$ emission and thin filament
\citep{b45} and molecular material \citep{b49} are also pointed
out.

\begin{figure}
 \includegraphics[width=8cm]{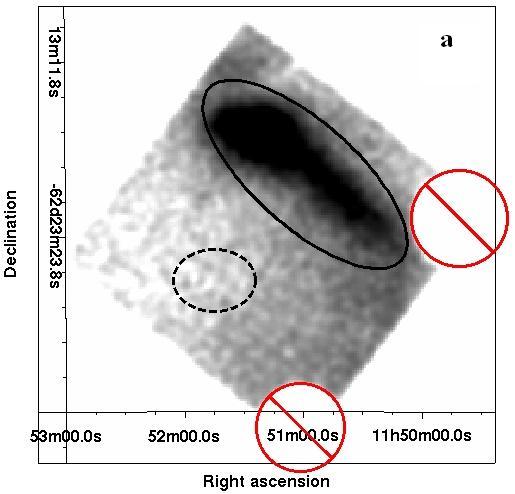}
 \includegraphics[width=8cm]{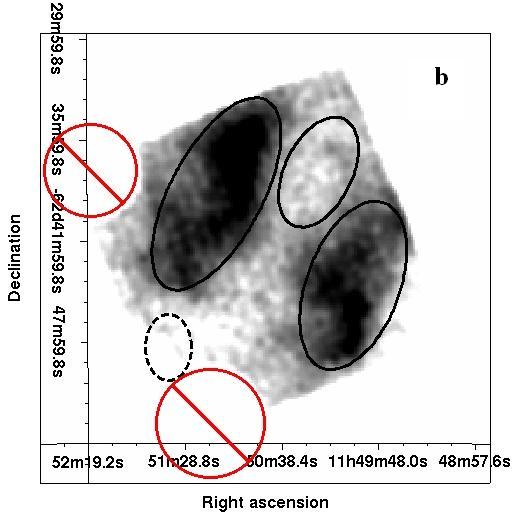}
  \caption{ {\it Suzaku} XIS0 image of (a) N region, (b) SE, SW and M regions of G296.1$-$0.5 in the 0.3$-$10 keV energy band (gray
scale). Regions for the source and the background are indicated
    with the solid and dashed ellipses, respectively. Two corners of each FOV, where the calibration
source of $^{55}$Fe is illuminated, are masked in these images.
The coordinates (RA and Dec.) are referred to epoch J2000.}
\end{figure}

\begin{figure}
 \includegraphics[width=8.5cm]{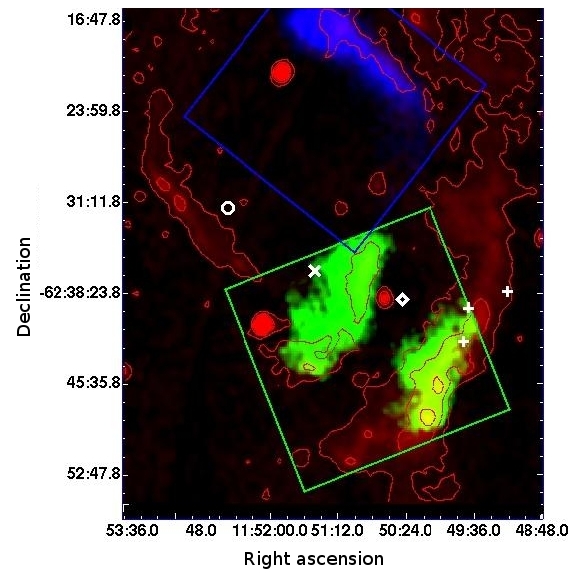}
  \caption{{\it Suzaku} X-ray mosaic image (blue for N and green for SE and
SW) and its radio image (red) obtained at 843
  MHz are overlaid and smoothed with
a Gaussian kernel of $\sigma$=2 arcsec, and units are in counts
s$^{-1}$. The linearly spaced radio contours are 0.02, 0.06, 0.10,
0.14, 0.18 and 0.22 counts/pixel. The blue and green squares
indicate the XIS FOV for N and S regions, respectively. The
positions of optical filaments \citep{b21}, H$\alpha$ emission,
thin filament \citep{b45} and molecular material \citep{b49} are
marked as, plus, circle, cross and diamond, respectively. The
coordinates (RA and Dec.) are referred to epoch J2000.}

\end{figure}

\subsection{Spectral Analysis}

We extracted the representative XIS spectra from four elliptical
regions named as  N, SE, SW and M (Fig. 1a and 1b) by using {\sc
xselect} version 2.4. The coordinates of the ellipses, their sizes
and angles are listed in Table 1. All spectra are grouped with a
minimum 20 counts bin$^{-1}$.

From the spectra we see that emission in 0.3$-$2.0 keV soft energy
range is dominant and K-shell lines of N, O, Ne, Mg, and Si are
present for all regions, while in 2$-$10 keV energy range the
emission is too faint. So, we select the energy range of 0.3$-$2.0
keV for spectral fitting.

To analyze XIS data of the selected regions, since previous {\it
XMM-Newton} observations showed non-equilibrium ionization (NEI)
condition in this SNR, we applied NEI collisional plasma with
variable abundances model (VNEI model in {\sc xspec}, \cite{b4})
modified by interstellar absorption (wabs model in {\sc xspec},
\cite{b18}) for all spectra. For N, SE and M regions,
one-component VNEI model best describes the spectral
characteristics of the emission. In the SW region, a two-component
VNEI model fits ($\chi^{2}$/d.o.f.=349.6/321) the emission better
than a one-component VNEI model ($\chi^{2}$/d.o.f.=403.6/324)
statistically. During the analysis, to see if there are small
scale variations the absorbing column density ($N_{\rm H}$),
electron temperature ($kT_{\rm e}$), the ionization timescale
($\tau$, which is often used as a key diagnostic of the NEI state,
is the product of the electron density multiplied by the time
after the shock heating) are allowed to vary. Since the emission
lines of N, O, Ne, Mg, and Si are clearly seen in the spectra,
their abundances are set as free parameters while the other
elemental abundances are fixed at their solar values \citep{b1}.
The best-fitting parameters, their errors and $\chi^{2}$ values
for all regions are given in Table 2 with corresponding errors at
the 90 per cent confidence limit. We note that, due to better
photon statistics of BI (XIS1) than FI (XIS0 and XIS3), only XIS1
results are given in this table. The background-subtracted XIS1
spectra in the 0.3$-$2.0 keV energy band for N, SE, SW and M
regions are shown in Figure 3. We also plot the confidence
contours of the absorbing column density ($N_{\rm H}$) versus the
electron temperature ($kT_{\rm e}$) and N abundance for all
regions in Figure 4. Both parameters are very well constrained for
N and SE regions.

\section{Discussion and Conclusions}

In this paper, we give a description of the X-ray emission from
G296.1$-$0.5 based on {\it Suzaku} archival data. We obtained a
clear image and high quality spectra of X-ray emission that lead
us detection of ejecta component in this middle-aged remnant.

The combined image, Fig. 2, shows that the remnant extends to a
large area (37$\times$25 arcmin$^{2}$) in radio band, while it is
concentrated in three regions as bright lobes in X-ray band, two
being in southern part and one being in the northern part. {\it
Suzaku} observed the lobes shown by squares, not the entire
remnant. Therefore we could study only the lobes and a region
between SE and SW lobes. The bright X-ray emission from lobes is
well correlated with radio emission. In both bands, G296.1$-$0.5
shows partial (irregular) shell morphology. Since radio emission
is not effected by absorbtion we can say that the shape of the
remnant is intrinsically anisotropic and it is interacting with a
non-uniform medium. The optical imaging performed by \citet{b21}
shows that SW region covers optical filaments with [S\,{\sc
ii}]/H${\alpha}$ ${\sim}$ 1.2 which is a typical value for shocked
gas. They obtained an electron density as high as $2\times10^{3}$
${\rm cm}^{-3}$. \citet{b45} detected strong localized diffuse
H${\alpha}$ emission using the data from the Marseille survey and
many thin filaments by equivalent AAO/UKST H${\alpha}$ survey in
the direction of the remnant. \citet{b49} detected molecular
material with a radial velocity of $\sim-38$ km s$^{-1}$ at
$\rmn{RA}(2000)=11^{\rmn{h}} 50^{\rmn{m}} 27^{\rmn{s}}$,
$\rmn{Dec.}~(2000)=-62\degr 38\arcmin55\arcsec$ by using CO
observations (see Fig. 2). All these observations may indicate
that a shock-cloud interaction is going on and the lobes may
result from these interactions.

X-ray emission of regions N, SE and M, can be described by
 one-component non-equilibrium ionization model (VNEI) in 0.3$-$2.0 keV band with
electron temperature $kT_{\rm e}$ in the range of ($0.51-0.76$)
keV which is in good agreement with the recent work by
\citet{b40}, while almost three times higher than that of
\citet{b39}. The ionization time scale, $\tau$, ranges between
$(1.25-2.46)\times10^{10}$ ${\rm cm^{-3}}$s indicating that the
plasma in these regions is far from ionization equilibrium
condition ($\tau$ is typically required to be $\geq$$10^{12}$ $\rm
cm^{-3}$s for full ionization equilibrium \citep {b14}). On the
other side, in the SW region, thermal emission can be adequately
described by two-component non-equilibrium ionization model
(VNEI+VNEI): the colder, $kT_{\rm e}\sim0.18$ keV, is
characterized by lower solar abundances of O and N and ionization
time scale $\tau\sim5.4\times10^{11}$ ${\rm cm^{-3}}$s, the
hotter, $kT_{\rm e}\sim0.84$ keV, by over abundances of Ne, Mg and
Si and ionization time scale $\tau\sim1.7\times10^{10}$ ${\rm
cm^{-3}}$s, both components are far from ionization equilibrium
condition in this region also.

The elemental abundance values of N, O, Ne, Mg and Si show small
variations in N, SE and M regions. In N and SE regions, the
abundance of O appears to be lower or about solar value, N, Ne, Mg
and Si values are somewhat higher than solar. In general, shocked
ISM spectra shows lower or about solar abundances. The plasma in
these two regions contains swept-up ISM and ejecta contamination.
In M region, abundance values and their errors seem somewhat
higher as compared to N and SE regions (see Table 2). We note that
the photon statistics in this region is poor (see Fig. 3),
therefore we can say that, like N and SE regions, the plasma
contains swept-up ISM and ejecta contamination. In SW region, the
abundances of N and O from cold component are lower solar
indicating ISM origin, while Ne, Mg and Si values from hot
component are high enough to show the evidence of ejecta origin in
this region also. G296.1$-$0.5 is a middle-aged remnant, thus the
presence of ejecta at its evolved stage is not expected. However,
recent X-ray observations show that in a number of middle-aged
remnants metal rich ejecta are still present (e.g. G156.2+5.7
\citep{b55}, N49B \citep{b44}, IC443 \citep{b43} and Puppis A
\citep{b57}). As seen in Fig. 2, the presence of molecular
material in the vicinity of SE and SW regions may cause the
survival of ejecta in especially these regions of this middle-aged
remnant.

The absorbing column density $N_{\rm H}$ value is ranging between
($2.0-14.6$)$\times10^{20}$ ${\rm cm^{-2}}$ across the remnant.
$N_{\rm H}$ value of SW region obtained from two temperature
model, $\sim$$14.6\times10^{20}$ ${\rm cm^{-2}}$, is in agreement
with the $N_{\rm H}$ value of \citet{b39} obtained by applying two
temperature model. On the other hand, $N_{\rm H}$ value of N and
SE regions are $2-3$ times smaller than $N_{\rm H}$ value of
\citet{b39} obtained by applying one temperature model, while
$4-5$ times higher than that of \citet{b40}. Therefore, for our
calculations we will use d=3 kpc value, that is in between the
values adopted by \citet{b39} and \citet{b40} and is in agreement
with the value estimated by \citet{b21}. We note that best-fitting
$N_{\rm H}$ value for SW region is relatively higher than the
other regions (see Table 2). The reason for this high $N_{\rm H}$
value may be the inhomogeneity of the foreground absorbing medium
(see Fig. 2). Since an accurate estimate of $N_{\rm H}$ put
constraints on the distance to the remnant, we obtained the
confidence contours of $N_{\rm H}$ versus $kT_{\rm e}$ and N
abundance as given in Fig. 4. We see that for N and SE regions
both parameters are very well constrained. The X-ray flux and the
corresponding luminosity values are obtained to be $F_{\rm
x}\sim1.1\times10^{-9}$ erg $\rm s^{-1}$$\rm cm^{-2}$ and $L_{\rm
x}\sim1.2\times10^{36}$ erg $\rm s^{-1}$ at d=3 kpc in the
0.3$-$2.0 keV energy range.

It is suggested that most optically observed SNRs are in the
adiabatic phase of their evolution \citep{b48}. Furthermore the
large physical size of this remnant, about 46 pc in radio, also
suggests that it is in adiabatic expansion phase. We calculated
physical parameters of G296.1$-$0.5 by using information obtained
by XIS images from Sedov equations \citep{b35}:

\begin{equation}
R_{\rm s} = 4\times10^{19}(\frac{t}{10^4 {\rm
yr}})^{2/5}(\frac{E}{10^{51} {\rm erg}})^{1/5} (\frac{n_0}{1
cm^{-3}})^{-1/5} {\rm cm},
\end{equation}

\begin{equation}
V_{\rm s}=5\times10^{7}(\frac{t}{10^4 {\rm
yr}})^{-3/5}(\frac{E}{10^{51} {\rm erg}})^{1/5} (\frac{n_0}{1
cm^{-3}})^{-1/5} {\rm cm s^{-1}},
\end{equation}

\begin{equation}
T_{\rm s}= 3\times10^{6}(\frac{t}{10^4 {\rm
yr}})^{-6/5}(\frac{E}{10^{51} {\rm erg}})^{2/5} (\frac{n_0}{1
cm^{-3}})^{-2/5} {\rm K},
\end{equation}

\begin{equation}
L_{\rm x}= 3.3\times10^{29}{n_0}^{2} {T_{\rm e}}^{0.5} {R_{\rm
s}}^{3} {\rm erg s^{-1}},
\end{equation}

\begin{equation}
M_{\rm ISM}= \frac{4}{3}\pi R_{\rm s}^{3}\mu m_{\rm H}n_{0},
\end{equation}

where $R_{\rm s}$ is the shock radius, $\sim$23 pc, the explosion
energy is assumed to be $E_{51}$=1, $\mu$ is atomic weight of
0.604, $m_{\rm H}$ is mass of a hydrogen atom. Through the
equations (1$-$5), we obtained ambient density $n_{0}\sim 0.45$
cm$^{-3}$, age of the remnant, $t\sim 2.8\times10^4$ yr,
confirming that G296.1$-$0.5 is a middle-aged SNR. The
corresponding ionization age $n_{\rm e}$t is then
$\sim$$2.7\times10^{11}$ ${\rm cm^{-3}}$s, a plasma with this
ionization age is in non-equilibrium ionization condition.
Assuming the electron-ion temperature equilibration, shock
velocity $V_{\rm s}$ that corresponds to best-fitting electron
temperature behind the main shock front ($kT_{\rm e}$ $\sim$ 0.18
keV) is calculated to be $\sim$320 km s$^{-1}$. Assuming uniform
density for the ambient ISM, we estimated shock temperature
$T_{\rm s}\sim 1.2\times10^{6}$ K and the mass swept-up by the
blast wave $M_{\rm sw}\sim 340$ {M\sun}. \citet{b40} suggested
that G296.1$-$0.5 is resulted from a core-collapse supernova with
a massive ($>25$ {M\sun}) progenitor star. We compared our
best-fitting abundances of N, O, Ne and Mg relative to Si for SE,
SW and M regions with abundance ratios in core-collapse model of
\citet{b60} and the obtained results are shown in Figure 5. As
seen from the figure, we can say that the progenitor star mass
might be $\ga25$ {M\sun}. Considering the high N abundance value
it appears to be roughly consistent with 30 {M\sun}. When we
compare this value with our estimated swept-up mass which is
approximately ten times higher, we can say that G296.1$-$0.5 is at
its critical stage and the contamination of ejecta in this
middle-aged SNR is not surprising.

\section*{Acknowledgments}

AS is supported by T\"{U}B\.{I}TAK PostDoctoral Fellowship. This
work is supported by the Akdeniz University Scientific Research
Project Management.

\begin{table*}
 \begin{minipage}{140mm}
  \caption{Centre coordinates, sizes and angles of N, SE, SW, M and background regions.}
 \begin{tabular}{@{}cccc@{}}
  \hline
     Regions & Centre Coordinates
& Size  & Angle \\
& $\rmn{RA}(2000)$ ; $\rmn{Dec.}~(2000)$
& (arcmin)  & (degrees) \\
& ($~^{\rmn{h}} ~^{\rmn{m}} ~^{\rmn{s}}$ ; $~\degr ~\arcmin
~\arcsec$)
&  &  \\
 \hline
N & 11 50 59 ; $-$62 19 46 & $7.5\times 3.0$ & $318$ \\
Background\footnote{for N region;} & 11 51 45 ; $-$62 25 56 & $2.4\times 1.8$ & $0$ \\
SE & 11 51 12 ; $-$62 39 13 & $2.7\times 6.3$ & $332$ \\
SW & 11 50 02 ; $-$62 44 38 &  $2.8\times 5.2$ & $339$ \\
M & 11 50 20 ; $-$62 37 55 & $1.9\times 3.5$ & $334$ \\
Background\footnote{for S region.} &11 51 37 ; $-$62 48 19 & $1.4\times1.9$ & $0$ \\
\hline
\end{tabular}
\end{minipage}
\end{table*}

\begin{table*}
 \begin{minipage}{180mm}
  \caption{Best spectral fitting parameters of XIS1 spectra obtained
for N, SE, SW and M region of G296.1$-$0.5 in 0.3$-$2.0 keV with
corresponding errors at 90 per cent confidence level.}
 \begin{tabular}{@{}cccccc@{}}
  \hline
     Component &Parameters & N & SE & SW & M \\
\hline
Absorbtion& $N_{\rm H}$($\times10^{20}$ $\rm cm^{-2}$) & 7.5 $\pm 0.5$& 4.7 $\pm 0.6$ & 14.6 $\pm 1.4$ & 2.0 $\pm 0.5$ \\
VNEI& $kT_{\rm e}$(keV) &0.58 $\pm 0.02$ & 0.51$\pm 0.02$ & 0.18$\pm 0.01$ & 0.76$\pm 0.15$ \\
Abundance\footnote{Abundance ratio relative to the solar value
\citep{b1}, (1) indicates that the elemental abundance is fixed at solar.}& N   & 1.91 $\pm 0.07$& 2.15 $\pm 0.09$ &  0.75 $\pm 0.12$ & 3.08 $\pm 0.36$  \\
& O   & 0.81 $\pm 0.02$& 1.01 $\pm 0.03$&  0.39 $\pm 0.03$ & 0.91 $\pm 0.07$   \\
& Ne  & 1.56 $\pm 0.04$& 1.92 $\pm 0.05$ & (1)& 2.3$\pm 0.2$  \\
& Mg  &  1.37$\pm 0.06$  & 1.44$\pm 0.10$ & (1) & 2.3$\pm 0.5$  \\
& Si  &   1.26$\pm 0.21$   & 1.75$\pm 0.36$ & (1) & 2.7$\pm 1.4$ \\
& $n_{\rm e}t$($\times10^{10}$ cm$^{-3}$s) & 2.24 $\pm 0.15$& 2.46$\pm 0.21$& 54.2$\pm 1.3$ &  1.25$\pm 0.29$  \\
& E.M.\footnote{Emission measure EM=$\int n_{\rm e}n_{\rm H}$dV in
the unit of $10^{57}$ $\rm cm^{-3}$, where $n_{\rm e}$ and $n_{\rm
H}$ are number densities of electrons and protons, respectively
and V is
the X-ray-emitting volume.} & 5.8$\pm 1.2$ & 4.6$\pm 1.1$ & 89.3$\pm 10.2$ & 0.59$\pm 0.03$  \\
VNEI&  $kT_{\rm e}$(keV) &  -  & -& 0.84$\pm 0.09$  &  - \\
&  Ne &  -  & -& 2.38$\pm 0.31$  &  - \\
&  Mg &   - & -& 1.55$\pm 0.21$  &  - \\
&  Si &  -  & -& 2.39$\pm 0.55$  &  - \\
&  $n_{\rm e}t$($\times10^{10}$ cm$^{-3}$s)  &  -  & -& 1.68$\pm 0.14$  & -  \\
&  E.M. &  -  &- & 1.8$\pm 0.4$&  - \\
& $\chi^{2}$/d.o.f.  & 619.2/380=1.63& 590.2/338=1.75 & 349.6/321=1.09 & 202.7/185=1.09  \\
\hline
\end{tabular}
\end{minipage}
\end{table*}

\onecolumn

\begin{figure}
 \centering
 \includegraphics[width=8.3cm]{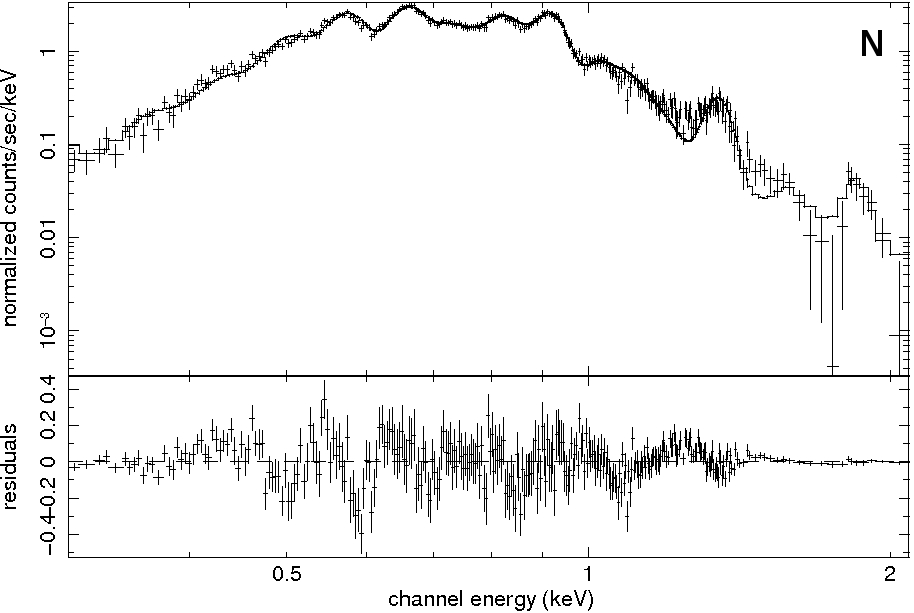}
\includegraphics[width=8.3cm]{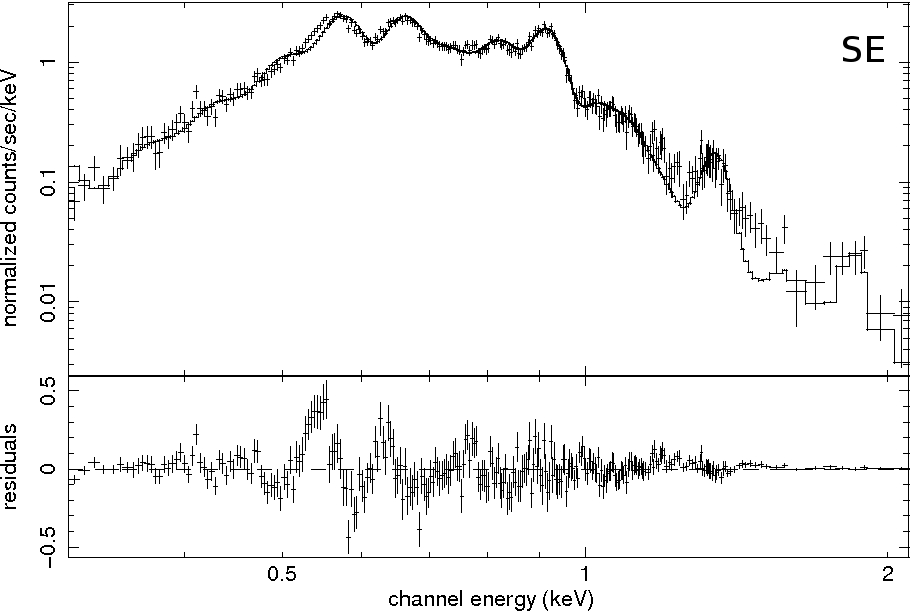}
\includegraphics[width=8.3cm]{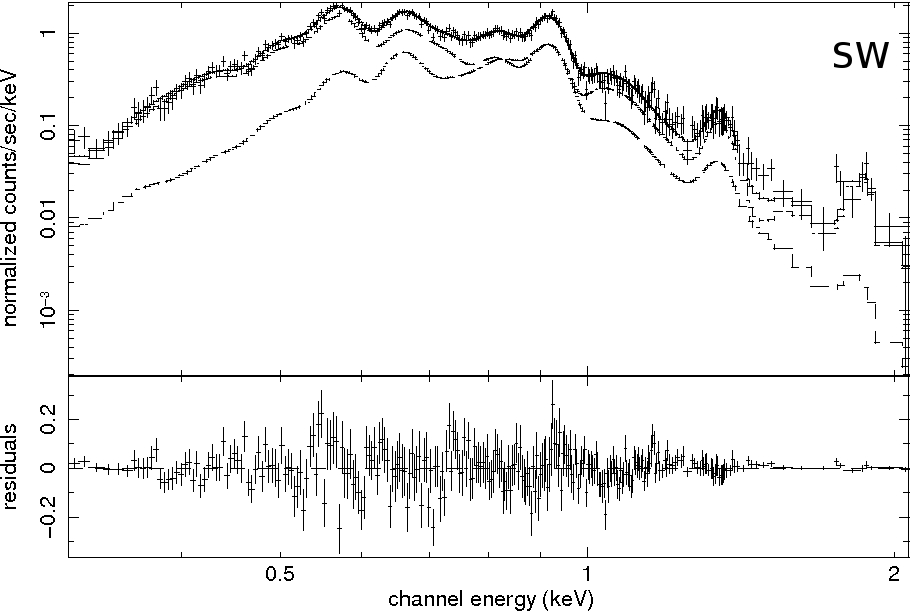}
\includegraphics[width=8.3cm]{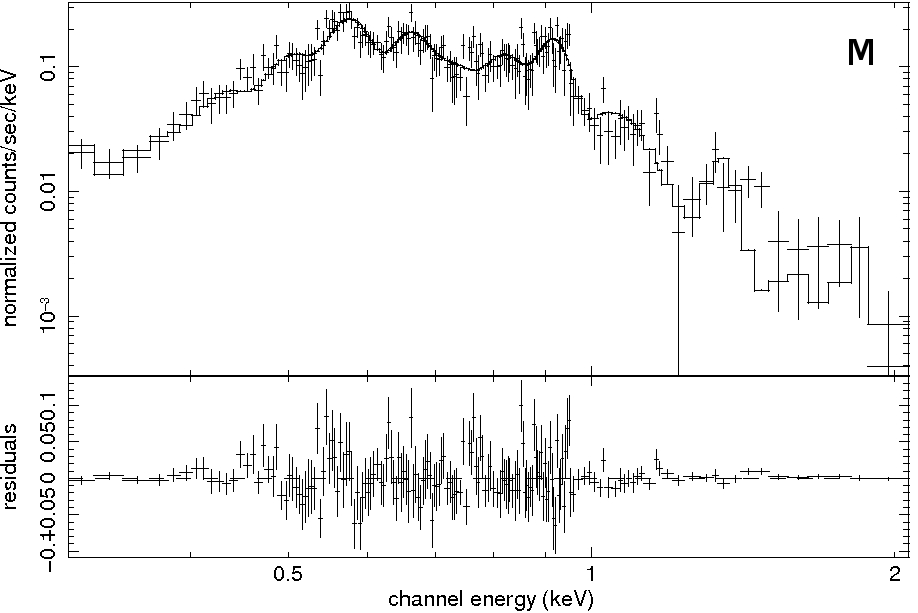}
\caption{XIS1 spectra of G296.1$-$0.5 obtained in the 0.3$-$2.0
keV energy band. The indicated fits consists of an absorbed VNEI
model for N, SE and M regions and an absorbed VNEI+VNEI model for
SW region. The bottom window gives the residuals from the
best-fitting model for XIS1 spectra.}
\end{figure}

\begin{figure}
 \centering
 \includegraphics[width=7cm]{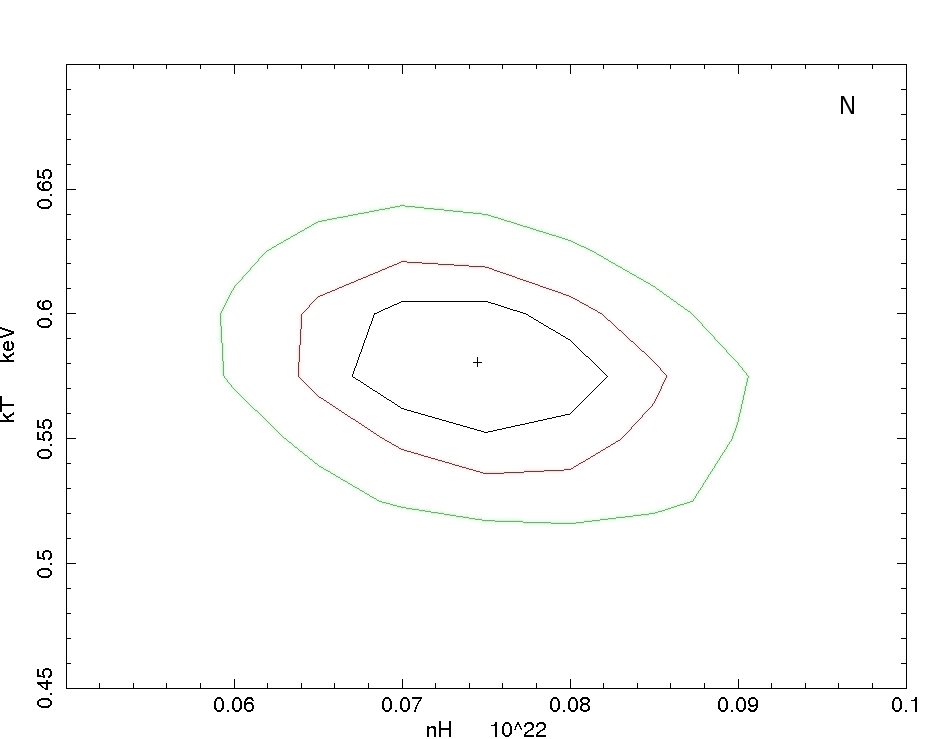}
\includegraphics[width=7cm]{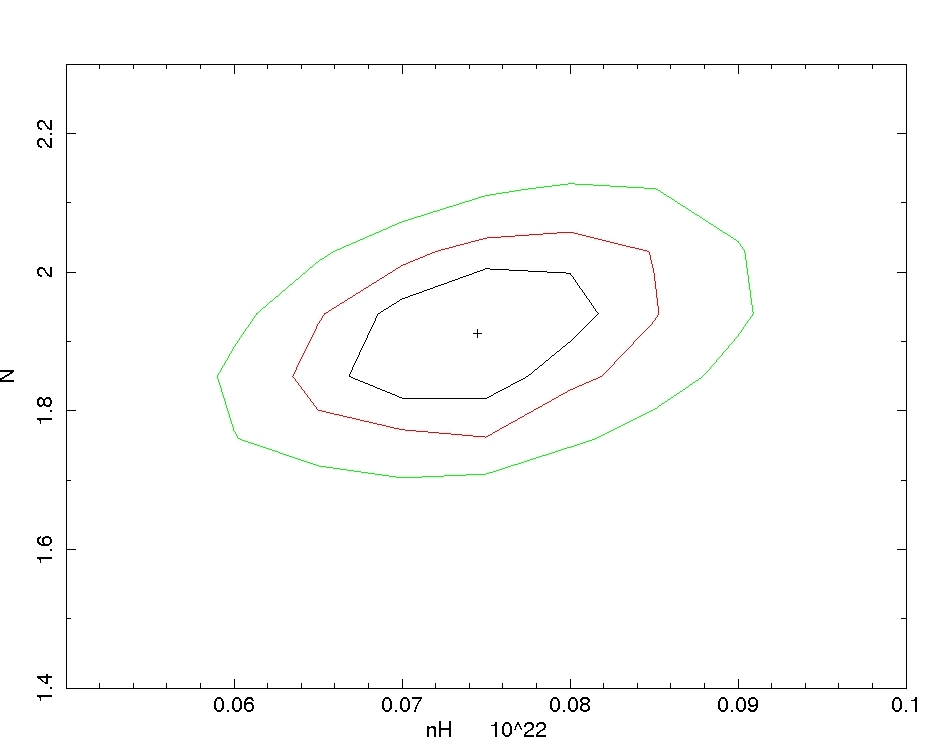}
\includegraphics[width=7cm]{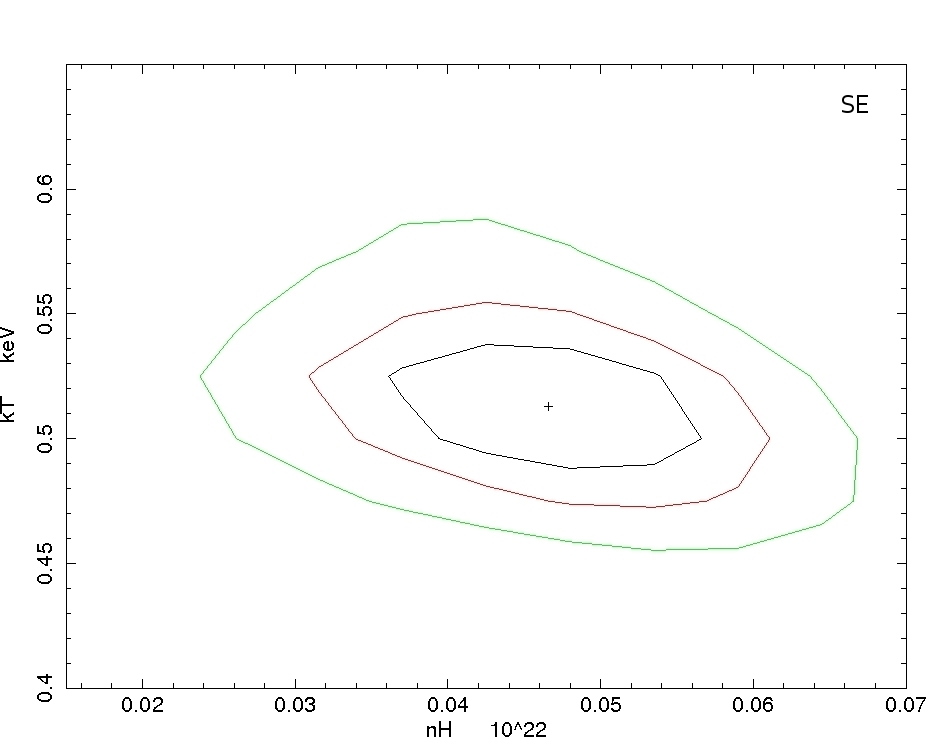}
\includegraphics[width=7cm]{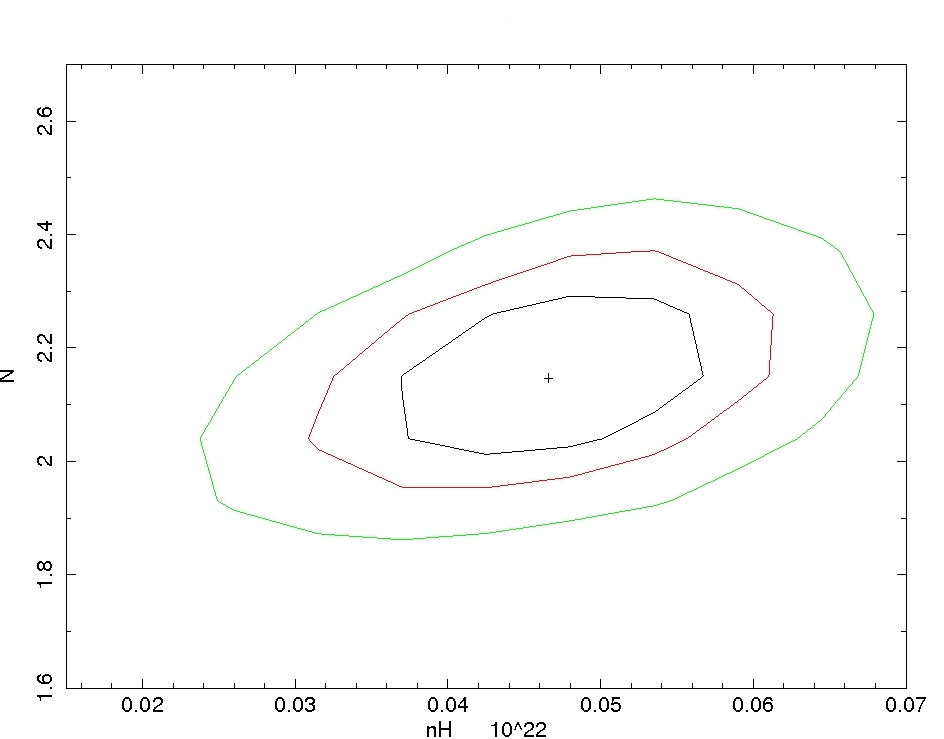}
\includegraphics[width=7cm]{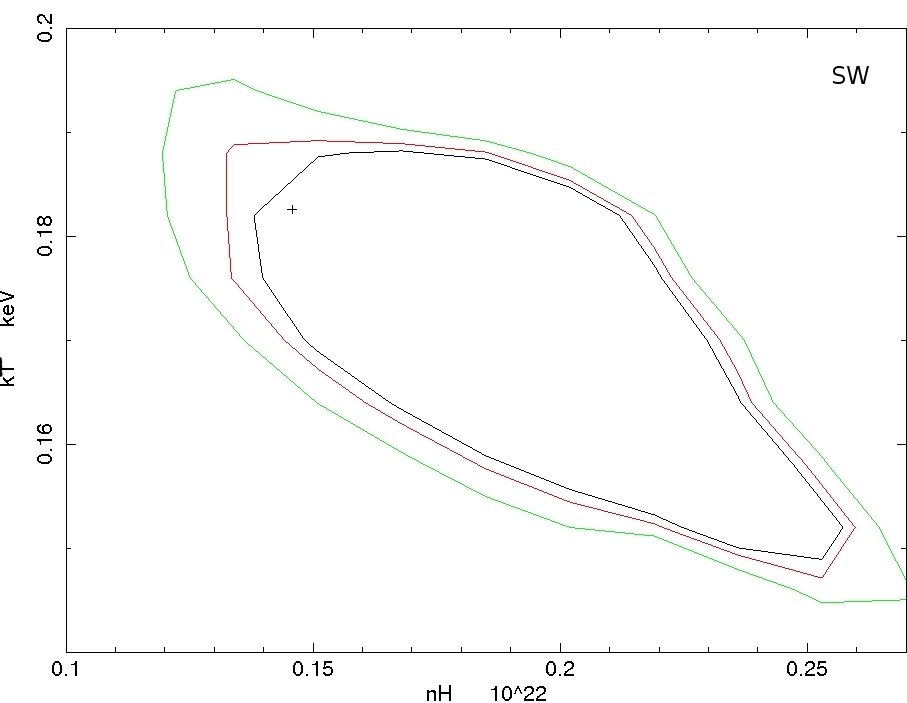}
\includegraphics[width=7cm]{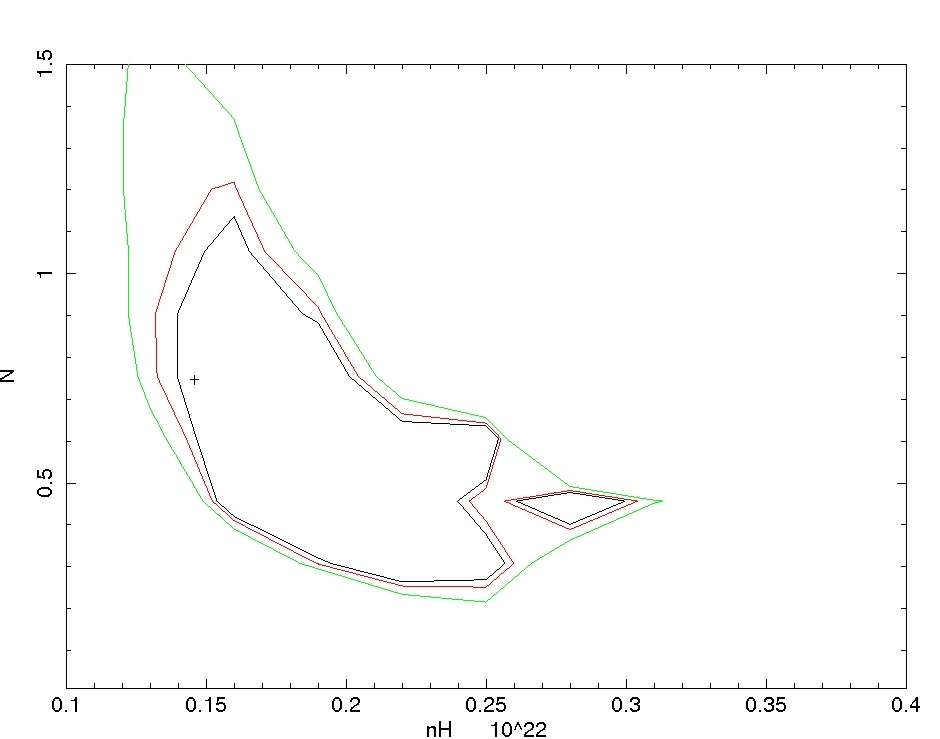}
\includegraphics[width=7cm]{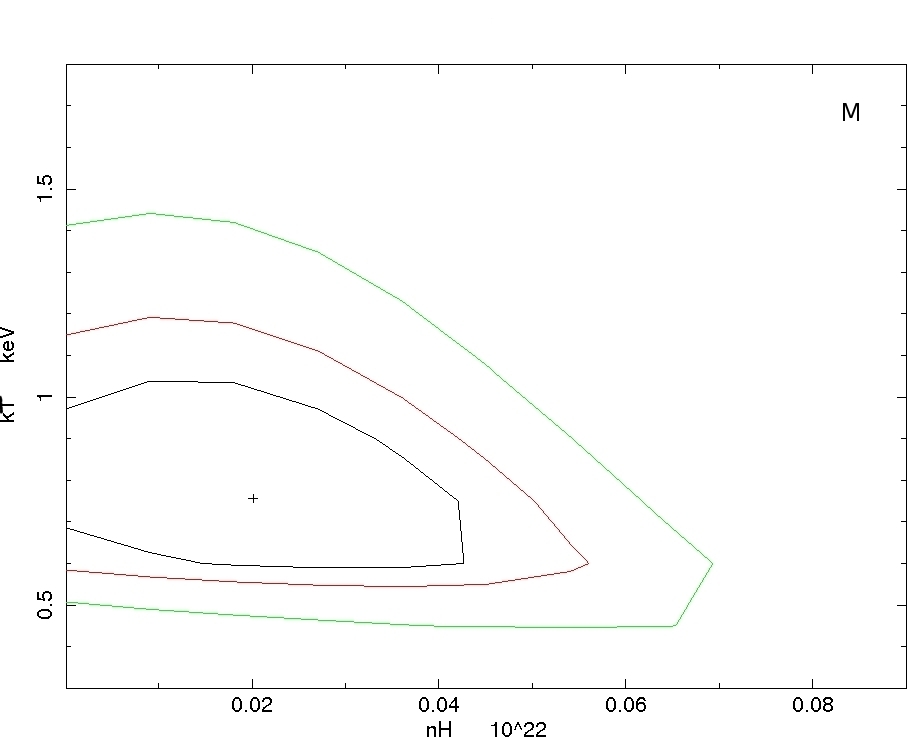}
\includegraphics[width=7cm]{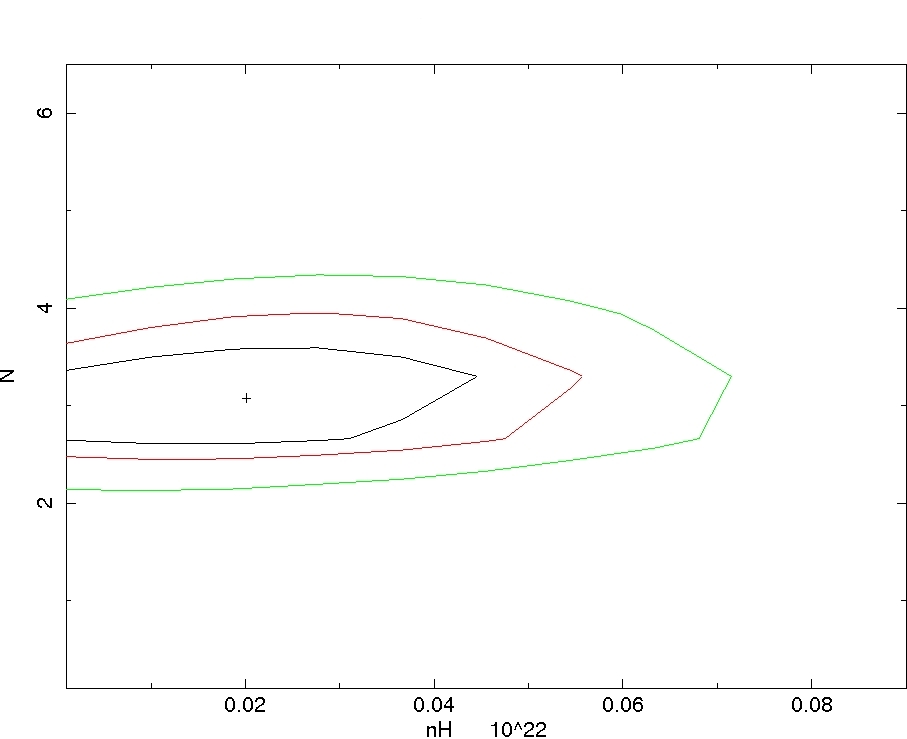}
\caption{Confidence contours of $N_{\rm H}$ versus $kT_{\rm e}$
(left) and $N_{\rm H}$ versus N abundance (right) spectral fitting
for all regions of G296.1$-$0.5. The contours are at the 68, 90
and 99 per cent confidence levels. The parameter values
corresponding to the best-fitting are marked by a plus sign (+).}
\end{figure}

\begin{figure}
 \centering
 \includegraphics[width=14cm]{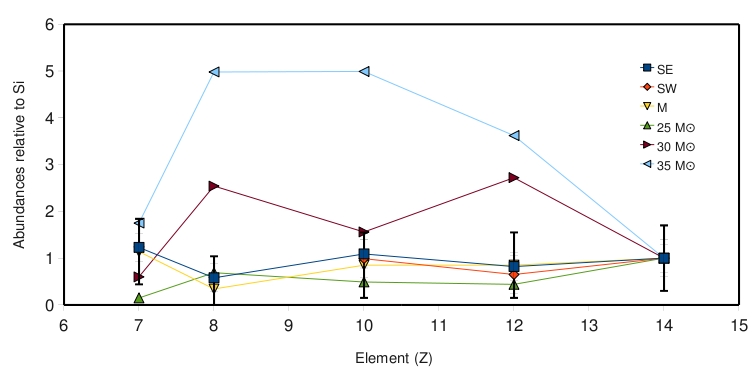}
\caption{Best-fitting abundances of N, O, Ne and Mg relative to Si
relative to solar (for SE, SW and M regions) are compared with the
theoretical predictions of core-collapse model \citep{b60} with
25, 30 and 35 {M\sun} progenitors. }
\end{figure}

\twocolumn

\end{document}